# Computing Economic Chaos

Richard H. Day
Department of Economics
University of Southern California
Los Angeles, CA 90089-0253

Oleg V. Pavlov
Information Systems Department
School of Management, Boston University
Boston, MA 02215

# Computing Economic Chaos[*]

Richard H. Day[**] and Oleg V. Pavlov[***]

July 2002

**Abstract**

Existence theory in economics is usually in real domains such as the findings of chaotic trajectories in models of economic growth, tâtonnement, or overlapping generations models. Computational examples, however, sometimes converge rapidly to cyclic orbits when in theory they should be nonperiodic almost surely. We explain this anomaly as the result of digital approximation and conclude that both theoretical and numerical behavior can still illuminate essential features of the real data.

## 1  Introduction

Mathematical existence theories in various scientific fields are usually developed in Euclidian or more general topological spaces. The mathematical entities whose "existences" are thereby established can not (in general) be represented by finite numbers. This fact is of particular relevance for economics where many variables such as prices or product quantities (cars, etc.) are naturally expressed as integers or rational fractions. Further, operational economic decisions invariably boil down to finite sequences of binary comparisons among discrete alternatives. This is the case whenever an optimization or equilibrium algorithm is used on a digital computer to solve a given problem numerically, as in the case of "computable general equilibrium models." It is also the case when a person solves a problem by "thinking it through." In economic problems, however, so long as they are formulated in finite dimensional Euchidian spaces, the mathematically existing, real valued solutions can usually be approximated as closely as one likes by using appropriate numerical techniques. In the rest of this paper such problems are called *computational*.

When a given real valued solution that is known to exist *cannot* be approximated, we shall say that it is *not computational*. The (perhaps clumsy) term "computational" is necessary to distinguish the issue we have in mind from the closely related but deeper theory of "computability" which would carry into more difficult issues than need to be dealt with here. Besides, that has already been done in the context of economics in a coruscating fashion by Velupillai (2000, forthcoming). The classical example of a *computational* problem is the circumference of a circle of diameter, say $D$. Even if $D$ is an integer or a rational fraction, the circumference, $C = \pi D$, cannot be computed exactly because $\pi$ is an irrational number. Yet, because $\pi$ can be computed to any desired degree of approximation—and has been to many thousands of places—the circumference is computational. The same can be said, of course, for squaring the circle.

However, there are theoretical economic problems that have been solved in real terms *that are not computational in the sense of this paper* and that is what we deal with here. This is the case of chaotic trajectories in dynamic economic processes. It is well know that many such processes, such as overlapping generations models, optimal and adaptive economizing growth models, tâtonnement, adaptive strategies in oligopolistic settings, and so forth, exhibit chaos and statistical (ergodic) behavior robustly.[1] The theorems that enable these properties to be established live in the reals. A special case has arisen in the economic models investigated by Nishimura and Yano (1995), Nishimura and Sorger (1996), and Hommes (1998). Although the models are different, involving, respectively, growth, business cycle and market mechanisms, in each case the trajectories are generated by iterations of the tent map.

In this paper we show why for *all* initial conditions, forward iterations simulationed on a binary computer using this map must converge to an *even, periodic orbit* even though for the same map defined over the reals, such convergence occurs with zero measure.[2] Nonetheless, it is possible to construct finite segments of trajectories that appear to be chaotic. Moreover, very small perturbations in the map can produce behavior that is highly irregular—so that it appears to be chaotic—and also produces histograms that approximate the real theoretical results.

We emphasize that the problem of computationality in our sense is well known. Anyone working with computational algorithms is bound to run into it sooner or later. For example, Stein and Ulam (1964, p.55) wrote about "spurious convergence" to cycles in computer experiments with the sine function, even though there are no attractive fixed points for the map in the real numbers. Farebrother (1991), Kaplan and Glass (1995) and Hommes (1998)—and no doubt many others—observed the same phenomenon. Their results can be duplicated with, for example, the freely available program *Dynamics* by Nusse and Yorke (1994). Still, students—even those who have developed sophisticated programming skills—are usually baffled by computational results that "contradict" the theory. Understanding this contradiction is crucial to guard against the tendency to reject one or the other of the contending outcomes for the wrong reason. In short, when the theory is not confirmed by the computer, both computer and real theory may be correct. It may just be that the latter is not computational!

Computer experimentation played a seminal role in the development of chaos theory by Lorenz, whose work motivated the Li–Yorke theorems that provided constructive conditions for establishing the theoretical existence and robustness of chaos in iterated maps. Thus, although the computer was helpful in discovering its existence, the results discussed here explain why chaos may sometimes not appear in computer experiments when supposedly it should.

## 2 The General Problem

Let $\theta : X \to X$ be a continuous map where $X$ is a closed, bounded interval in the non–negative, real numbers, $R$. Such a map defines a discrete time dynamic process

$$x_{t+1} = \theta(x_t), \ t = 0, 1, \ldots \quad \text{where} \quad x_0 = x \in X. \tag{1}$$

The trajectory of such a map is defined by $\tau(x) := \{\theta^t(x)\}_{t=0}^{\infty}$ where $\theta^0(x) := x_0$ and $\theta^1(x) := \theta(x)$ and $\theta^t(x) = \theta(\theta^{t-1}(x))$. Thus,

$$x_t = \theta^t(x_0), \ t = 0, 1, 2, \ldots. \tag{2}$$

Li and Yorke (1975a, 1975b) derived simple, constructive criteria to show that such a map could generate chaotic trajectories (in $R$) and provided conditions that the relative frequency distribution of values in any such trajectory converged to a unique, absolute, continuous, invariant measure or density function. Given both conditions, chaos "occurs almost surely" for any initial condition drawn at random in $X$.

To pin down exactly what is at the heart of the issue, a few basic properties of binary arithmetic need to be reviewed.[3] First, every number is represented by a sequence of zeros and ones, the integer portion by a string of $p$ zeros or ones and the fractional part by a string of $q$ zeros or ones where $p$ and $q$ are positive integers. Let $m$ be the maximum number of significant digits that can be represented, a number determined by the word length in the computer and the precision to be used (single, double, etc.).[4] The cardinal of the set of numbers, $\bar{X}$, representable on the computer is $2^m$ where $p + q = m$.

Consider any element $x$ in $X$. Define a *round–off map*, $\rho : X \to \bar{X}$ by[5]

$$\rho(x) = \text{ the binary number given by the first } m \text{ significant digits of } X.$$

Returning to the domain of our dynamic process, any $x \in X$ becomes the number $\bar{x} = \rho(x)$ on the computer.

$$\bar{X} = \rho(X) := \left\{ \bar{x} = \rho(x) \quad \text{for all} \quad x \in X \right\}$$

is the set of $2^m$ rational numbers in $X$ with at most $m$ significant digits. Obviously, any initial condition that must be truncated introduces a round–off error at the outset.

For most nonlinear maps the value of $\theta(x)$ must be approximated by an algorithm that terminates after a finite number of steps with an element in $X$. On the computer such an algorithm in effect defines a map $\bar{\theta} : \bar{X} \to X$ such that for all $x \in \bar{X}$, $\bar{\theta}(\bar{x}) \in X$. Whenever this occurs, a new round–off error is introduced in addition to those that might arise in the algorithmic steps. The map $(\rho \cdot \bar{\theta})(x)$, therefore, is the computational representation of equation (1) and the *computer analog of the real dynamical process* is

$$\bar{x}_{t+1} = (\rho \cdot \bar{\theta})(\bar{x}_t), \ t = 0, 1, \ldots \quad \text{where} \quad \bar{x}_0 = \rho(x). \tag{3}$$

The sequence of computational iterated maps is

$$\bar{x}_t = (\rho \cdot \bar{\theta})^t \bar{x}_0 \quad \text{where} \quad \bar{x}_0 = \rho(x_0). \tag{4}$$

By construction $\rho \cdot \bar{\theta}$ maps $\rho(X)$ into $\rho(X)$. The immediate implication is that

**A.** *Every computed trajectory is cyclic or converges to a cycle in finite time.*[6]

The general computational problem is that the repeated truncation in a given simulation can accumulate error so that a computed sequence, $\bar{x}_0, \ldots$ can be far removed from its theoretical counterpart, $x_0, \ldots$. That is,

$$E_T = \sum_{t=0}^{T} \left| (\rho \cdot \bar{\theta})^t(\rho(x)) - \theta^t(x) \right| \tag{5}$$

can grow without bound as $T$ becomes large. This is exactly what happens in the theoretical examples at issue where the real trajectories generated by $\theta$ are chaotic, and those generated by $\rho \cdot \bar{\theta}$ are cyclic.

This is almost all there is to the story except to show that the cycles are *even* in the case of the tent map.

## 3  The Tent Map

The family of tent maps that underlie the economic models cited above is

$$\theta(x) := \begin{cases} ax & \text{if } 0 \leq x < \frac{N}{2} \\ a(N-x) & \text{if } \frac{N}{2} \leq x \leq N. \end{cases} \tag{6}$$

For convenience, we consider examples where $a = 2$ and $N$ is a positive integer. Points that satisfy the Li–Yorke (1975) overshoot conditions that imply the existence of chaotic trajectories and periodic orbits of every periodicity are easily found. It is also relatively easy to show that the uniform distribution on $[0, N]$ is the unique absolutely continuous invariant measure for the map $\theta : [0, N] \to [0, N]$. Those facts imply that (in the sense of the uniform measure) for almost all $x \in [0, N]$ the real trajectories are nonperiodic and chaotic. See, for example, Day (1994, p.86). Here, then is an example of existence theory in the reals.

The first iterate of the tent map is just equation (6). The second iterate is given by

$$\theta^2(x) = \begin{cases} 4x & \text{if } 0 \leq x < \tfrac{1}{4}N \\ 4(\tfrac{2}{4}N - x) & \text{if } \tfrac{1}{4}N \leq x < \tfrac{2}{4}N \\ 4(x - \tfrac{2}{4}N) & \text{if } \tfrac{2}{4}N \leq x < \tfrac{3}{4}N \\ 4(N - x) & \text{if } \tfrac{3}{4}N \leq x \leq N \end{cases}$$

the third by

$$\theta^3(x) = \begin{cases} 8x & \text{if } 0 \leq x < \tfrac{N}{8} \\ 8(\tfrac{N}{4} - x) & \text{if } \tfrac{N}{8} \leq x < \tfrac{N}{4} \\ 8(x - \tfrac{N}{4} - x) & \text{if } \tfrac{N}{4} \leq x < \tfrac{3}{8}N \\ 8(\tfrac{N}{2} - x) & \text{if } \tfrac{3}{8}N \leq x < \tfrac{N}{2} \\ 8(x - \tfrac{N}{2}) & \text{if } \tfrac{N}{2} \leq x < \tfrac{5}{8}N \\ 8(\tfrac{3}{4}N - x) & \text{if } \tfrac{5}{8}N \leq x < \tfrac{3}{4}N \\ 8(x - \tfrac{3}{4}N) & \text{if } \tfrac{3}{4}N \leq x < \tfrac{7}{8}N \\ 8(N - x) & \text{if } \tfrac{7}{8}N \leq x \leq N. \end{cases}$$

The $n^{\text{th}}$ iterate can be expressed by

$$\theta^n(x) = \begin{cases} 2^n x & \text{if } 0 \leq x < \tfrac{N}{2^n}, \\ 2^n(\tfrac{2iN}{2^n} - x) & \text{if } \tfrac{N}{2^n}(2i-1) \leq x < \tfrac{N}{2^n}2i,\ i = 1, \ldots, 2^{n-1}, \\ 2^n(x - \tfrac{2iN}{2^n}) & \text{if } \tfrac{N}{2^n}2i \leq x < \tfrac{N}{2^n}(2i+1),\ i = 1, \ldots, 2^{n-1} - 1. \end{cases} \quad (7)$$

Notice that as $a$ and $N$ are integers, an algorithmic approximation is not needed. That is, $\theta \equiv \bar{\theta}$.

Nonetheless, round–off error must appear. Equation (7) indicates that the $n^{\text{th}}$ iterate of $x$—regardless in which interval the initial condition lies—contains the term $2^n x$. Clearly, the fractional part of the initial condition $x$ becomes shorter by one element with each multiplication by 2.[7] Therefore, *given the precision $m$, $2^n x$ must become an integer for some $n \leq m$*. The terms $2iN$ are even integers. The other term is always an integer. The difference of two integers is an integer and any integer multiplied by $2^n$ is even, so after some number of iterations, not more than $m$, $\theta^t(x)$ becomes an even integer.

**B.** *If $a = 2$ and $N$ is an integer, then, after a finite number of iterations, every computed trajectory of (6) will have only even integer values.*

From **A** every trajectory converges to a cycle in finite time. From **B** every trajectory converges to even integers. Therefore, we have

**C.** *For $a = 2$ and any integer $N$ and given finite precision $m$, every computed trajectory of the tent map (6) converges in finite time to a periodic orbit consisting of even integers.*

Thus, viewing the computer itself as a dynamical system, the chaos that occurs almost surely in the reals surely does not occur in the binary computer.

If $N = 1$, then $x_0$ is always a fraction, zero, or one. Zero is always a fixed point and $\theta(1) = 0$. Since the fractional part goes to zero, we have the following corollary.

**D.** *If $N = 1$, every computed trajectory converges to zero in finite time.*

This is the result obtained numerically by Kaplan and Glass (1995) and observed by Hommes.

## 4    Numerical Examples

To better understand the above corollary, let us look at an example. For simplicity, assume that we work on a computer that stores binary numbers as a string of five binary digits and that the computer does not use binary normalization, that is a binary number can start with a zero. For the initial decimal value of $x_0 = 0.4$, equation (1) for $N = 1$ produces the periodic sequence: 0.4, 0.8, 0.4, 0.8, 0.4,.... However, due to the round off error in representing the initial value and due to the disappearance of the fractional part the numerical sequence will converge to zero, as **D** predicts.[8] Rewrite equation (6) for $N = 1$ using binary numbers in place of the decimal numbers:

$$\theta(x) := \begin{cases} (10.0)_2 \times x & \text{if } (0.0)_2 \leq x < (0.1)_2 \\ (10.0)_2 \times ((1.0)_2 - x) & \text{if } (0.1)_2 \leq x \leq (1.0)_2 \end{cases} \quad (8)$$

Iterating this binary representation of the tent map on our imaginary computer would produce a binary sequence that is shown below together with its decimal equivalents.

| operation (in binary numbers) | binary number | decimal equivalent |
|---|---|---|
| binary representation of $x_0 = 0.4$ | 0.0110 | $0 \times \frac{1}{2} + 1 \times \frac{1}{2^2} + 1 \times \frac{1}{2^3} + 0 \times \frac{1}{2^4} = 0.375$ |
| $0.0110 < 0.1 \Rightarrow 10.0 \times 0.0110$ | 0.1100 | $1 \times \frac{1}{2} + 1 \times \frac{1}{2^2} + 0 \times \frac{1}{2^3} + 0 \times \frac{1}{2^4} = 0.75$ |
| $0.1100 > 0.1 \Rightarrow 10.0 \times (1.0 - 0.1100)$ | 0.0100 | $0 \times \frac{1}{2} + 1 \times \frac{1}{2^2} + 0 \times \frac{1}{2^3} + 0 \times \frac{1}{2^4} = 0.25$ |
| $0.0100 < 0.1 \Rightarrow 10.0 \times 0.0100$ | 0.1000 | $1 \times \frac{1}{2} + 0 \times \frac{1}{2^2} + 0 \times \frac{1}{2^3} + 0 \times \frac{1}{2^4} = 0.5$ |
| $0.1000 > 0.1 \Rightarrow 10.0 \times (1.0 - 0.1000)$ | 1.0000 | $1 \times 2^0 = 1$ |
| $1.0000 > 0.1 \Rightarrow 10.0 \times (1.0 - 1.0000)$ | 0.0000 | $0 \times \frac{1}{2} + 0 \times \frac{1}{2^2} + 0 \times \frac{1}{2^3} + 0 \times \frac{1}{2^4} = 0.$ |

Notice that the numerical binary sequence is not equivalent to the periodic decimal sequence. Thus, instead of the 2–period orbit, $\{.4, .8\}$, we get the sequence $.375, .75, .5, 1, 0$ converging to the stationary state zero.

Consider what happens when $N = 100$. We computed trajectories using a C++ program that was compiled with the *Borland C++* compiler installed on a Pentium computer. It took only 16 iterations for the trajectory starting at $x_0 = 67.2$ to converge to the integer 19 that then led to the 10-period orbit $\{8, 16, 24, 32, 48, 56, 64, 72, 88, 96\}$. Figure 1 displays this cycle with its integer preimages. The lowest level of the tree contains odd integers. For the initial value $x_0 = 4.23828125$ the trajectories converged to the integer 85 in 8 iterations, leading to the 2–period orbit $\{40, 80\}$. Figure 2 gives this orbit with its integer preimages.

Figure 1: Ten period cycle and its integer preimages

Figure 2: Two period cycle and its integer preimages

Figure 3: Fixed point and its integer preimages

Of course, the empirical densities for these orbits are just spikes over the cyclic points and not integrable. For the initial condition $x_0 = 12.5$ the trajectory quickly converged to the fixed point of 0. See Figure 3. The integer preimages of the three orbits and the cyclic points themselves account for all the integers in $[0, N]$. Note that the elements of all three orbits are even.

## 5  Imitating Chaos

In spite of these results it is possible to approximate chaos. We consider two possibilities.

Figure 4: Histogram of Preimages for $N = 100$, $x_0 = 67.2$

## 5.1 Constructing Preimages

Any point $x$ in $[0, N]$ has two preimages, each of which has two and so on of the difference equation defined by equation (6). At each point we can choose one of the two preimages, find its two preimages, choose one, and so on. This can be done for some number say $p$ times starting for any arbitrary number $\bar{x} \in \rho\big([0, N]\big)$ in this way generating a sequence $x_0, x_1, \ldots, x_p$. If the choice at each stage is "random" (say, by tossing a coin or using a computer random number generator), this sequence will look more or less chaotic. Moreover, the sequence $x_{-p}, x_{-p+1}, \ldots, x_0$ will be a segment (possibly approximate) of the real trajectory of (1) starting at the initial condition $x_{-p}$. However, if we begin computing the difference equation (3) from that same point, $(x_{-p})$, we will not get the irregular segment just constructed, but will quickly converge to an even cycle instead!

For example, we took the initial condition 67.2 and constructed in the above manner 60,000 preimages to obtain the numerical density shown in Figure 4. It roughly approximates the uniform density, suggesting that the absolutely continuous invariant *density* of the real trajectories is computational.

In spite of this, "chaotic" trajectories so constructed are not computational. The 60,000$^{\text{th}}$ preimage has a fractional part. Consequently, *forward computation from that point cannot follow the true constructed trajectory but will quickly converge to an even integer cycle as predicted by the theorem* and shown in our examples of section 5.

Figure 5: Histogram for $N = 100.0001$, $x_0 = 67.2$

While studying convergence to cycles for the map $y_{t+1} = \sin \pi y_t$, $0 \leq y \leq 1$, Stein and Ulam suggested inverse iterations with random choice between left and right preimages. They point out that for many other maps, the probability of convergence to low order cycles is very small. Even for the tent map, however, a good approximation to chaos can be attained in another way.

## 5.2 Noninteger domain

By setting $N$ to a non–integer value, we can keep $x$ from converging quickly to an integer. For example, choosing $N = 100.0001$, rather than 100, as in the previous experiment, gives us the distribution of Figure 5 ($x_0 = 67.2$, $60,000$ iterations). It is "close" to the uniform distribution that is predicted by the theory for the tent map over the reals; and it resembles the distribution obtained using randomly selected backward iterates.

# 6  Discussion

In spite of the anomalies discussed above, experience shows that for most maps forward iterations generate trajectories that on a computer *appear* to be chaotic when in theory they "should be:" irregular fluctuations are generated and they seem to obey the usual statistical laws of large numbers, even though ultimately the *computations* must be cyclic. Thus, the applied models *can* provide potential partial explanations for the ubiquitous irregularity of fluctuation in economic data.[9] If we acknowledge that much (if not all) economic data arise in the finitely representable rationals (as we must), then theoretical results derived in the reals must be conceptual approximations of the real world economic phenomena. Thus, it seems to us that the "anomaly" analyzed here does not vitiate inferences based on the real theoretical properties of a given economic model—at least not from a practical point of view. Real theory can still lead us to a better understanding of actual economic data—most of which is, indeed, highly irregular. Of course, the theory *is* an idealization. But that is a general characteristic of theory, the purpose of which, after all, *is* to idealize, that is, to help us understand in the simplified terms demanded by logic the unfathomably complex world around us. Less appreciated is the possibility that a numerical analog of a theory could be entirely misleading, less appreciated, perhaps, because it seems not to occur often in practice.

## 7 Appendix: Binary Numbers

We present here a brief synopsis of the binary number system. Every real number, $x \in R$, consists of integer and fractional parts and can be uniquely represented in a base, $r$, by an infinite sequence,

$$
\begin{aligned}
x_r &= (a_p a_{p-1}, \ldots, a_1 a_0 \cdot a_1 a_2, \ldots, a_q, \ldots)_r \\
&:= \left[(a_p \times r^p) + (a_{p-1} \times r^{p-1}) + \cdots + (a_1 \times r^1) + (a_0 \times r^0)\right] + \\
&\quad \left[(a_{-1} \times \tfrac{1}{r}) + \cdots + (a_{-q} \times \tfrac{1}{r^q}) + \cdots\right]
\end{aligned}
\tag{9}
$$

where $a_i \in \{0, \ldots, r-1\}$. When $r = 2$ and we have the binary numbers $D$, $a_i \in \{0, 1\}$, $i \in \{p, p-1, \ldots, 1, 0, -1, -2, \ldots, -q\}$. When $r = 10$, we have the commonly used decimal numbers.

The decimal number

$$(67.2)_{10} = 6 \cdot 10^1 + 7 \cdot 10^0 + 2 \cdot \left(\frac{1}{10}\right)^1, \tag{10}$$

whereas,

$$(0.4)_{10} = 4 \cdot \left(\frac{1}{10}\right)^1. \tag{11}$$

Notice that in both cases the decimal numbers have finite representations.

A binary computer converts decimal numbers to binary representation. The algorithm for doing this is given below. In the preceding examples,

$$
\begin{aligned}
(67.2)_{10} &= 2^6 + 2^1 + 2^0 + \left(\tfrac{1}{2}\right)^3 + \left(\tfrac{1}{2}\right)^4 + \left(\tfrac{1}{2}\right)^7 + \left(\tfrac{1}{2}\right)^8 + \ldots \\
&= (1000011.0011\overline{0011})_2 \\
&\simeq 1000011.00110011001100110\ldots.
\end{aligned}
\qquad (12)
$$

$$
\begin{aligned}
(0.4)_{10} &= 0 + \left(\tfrac{1}{2}\right)^2 + \left(\tfrac{1}{2}\right)^3 + 0 + 0 + \left(\tfrac{1}{2}\right)^6 + \left(\tfrac{1}{2}\right)^7 + 0 + \ldots \\
&= (0.\overline{0110})_2 \\
&\simeq 0.011001100110\ldots.
\end{aligned}
\qquad (13)
$$

In each case the binary representation involves an infinite series. Consequently, *even though they are rational, the numbers must be approximated on any computer with finite precision.*

**Binary Arithmetic and Computer Representation**

Consider a decimal number with integer and fractional parts such as 67.4 as shown in section 5.1. To translate the integer part we divide it by 2, keeping the remainders as binary digits and adding 1 as the most significant digit:

|  | **Remainder** |  |  |
|---|---|---|---|
| $67 \div 2 = 33$ | 1 | => | $a_0 = 1$ (least significant digit) |
| $33 \div 2 = 16$ | 1 | => | $a_1 = 1$ |
| $16 \div 2 = 8$ | 0 | => | $a_2 = 0$ |
| $8 \div 2 = 4$ | 0 | => | $a_3 = 0$ |
| $4 \div 2 = 2$ | 0 | => | $a_4 = 0$ |
| $2 \div 2 = 0$ | 0 | => | $a_5 = 0$ |
|  |  |  | $a_6 = 1$ (most significant digit) |

The decimal part is multiplied by 2:

$$
\begin{aligned}
0.2 \times 2 &= 0.4 &&=> \quad a_{-1} = 0 \\
0.4 \times 2 &= 0.8 &&=> \quad a_{-2} = 0 \\
0.8 \times 2 &= 1.6 &&=> \quad a_{-3} = 1 \\
0.6 \times 2 &= 1.2 &&=> \quad a_{-4} = 1 \\
0.2 \times 2 &= 0.4 &&=> \quad a_{-5} = 0 \\
&\ldots
\end{aligned}
$$

Then we put the integer and fractional parts together:
$$(67.2)_{10} = \left(1000011.0011\overline{0011}\right)_2.$$

## Notes

*The authors gratefully acknowledge comments of Amy Radunskya and two referees.

**Department of Economics, University of Southern California, Los Angeles, CA 90089-0253.

***Information Systems Department, School of Management, Boston University, Boston, MA 02215.

**1**. Many examples are given, for example, in Grandmont (1987) or Day (1994, 1999).

**2**. The concepts of measure theory are fundamental to the present discussion for they are necessary to characterize the statistical behavior of chaotic trajectories and to specify whether or not real chaos is robust with respect to initial conditions in the reals. This is the case we investigate. That is why the innocent might be astonished not to find it on the computer. For the far more difficult issue of extending the concept of measure to sequences of numbers that may not converge to cycles, see Kolmogorous (1998) and Martin–Löf (1966).

**3**. See the appendix in §7 for examples.

**4**. Precision is determined by the number of significant digits that can be carried. For example, 1.0000 and 1.0001 have five significant digits each, while .0001 has a single significant digit. See Lipschultz (pp.59, 60) for a definition of "precision."

**5**. The mere conversion of a number to a different base can introduce round–off. See section 7, equations (10)–(13) for examples. For a review of these topics, see, for example, Aho, et al. (1992). The importance of round off error in digital computations has recently received close attention in an entirely different context; that of computational algorithms used in econometric estimation. See McCullough and Vinod (1999). Such problems were recognized decades ago by Goldberger and Zellner when their students, using different regression packages, arrived at markedly different parameters in linear regressions.

**6**. Proof: Pick a point $x_0 \in \bar{X}$ and compute a sequence $(\rho \cdot \bar{\theta})(x_0), (\rho \cdot \bar{\theta})^2(x_0), \ldots, (\rho \cdot \bar{\theta})^t(x_0)$ such that $(\rho \cdot \bar{\theta})^t(x_0) = (\rho \cdot \bar{\theta})^s(x_0)$ for some $s$, $1 \leq \underline{x} \leq t \leq m-1$. Such a point must exist for otherwise $(\rho \cdot \bar{\theta})^m(x_0) \notin \bar{X}$ contrary to hypothesis. This implies that points $(\rho \cdot \bar{\theta})^s(x_0), \ldots, (\rho \cdot \bar{\theta})^{t-1}(x_0)$ is a $t-s$ cycle. Let $S_i = \{x_0^1, (\rho \cdot \bar{\theta})(x_0^1), \ldots, (\rho \cdot \bar{\theta})^{t-1}(x_0^1)\}$ and add to this set all the preimages of $x_0^1$ in $\bar{X}$. All of these end in the $t-s$ cycle just constructed. Include them in $S_1$. If $S_1 = \bar{X}$, we are done. If not, pick a point in $\bar{X} \setminus S_1$ and repeat the process, in this way constructing a finite sequence of sets $S_1, \ldots, S_n, k \geq 1$, such that

$$\cup_i S_i = \bar{X},$$

and such that every trajectory beginning in $S_i$ ends in a stationary point or a finite cycle.

**7**. Everyone knows that multiplying a decimal number by 10 can be done merely by shifting the decimal one place to the right. Dividing by 10 involves shifting the decimal one space to the left. The situation is analogous for the binaries. Multiplying a binary number by 2 is equivalent to shifting the binary point one digit to the right. Conversely, division by 2 shifts the binary point one place to the left. It is this fact that means that the tent map (6) has an exact binary digital representation for $a = 2$.

**8**. The exact binary representation of the decimal 0.4 is $(0.\overline{0110})_2$. The bar implies an infinite repetition of the sequence 0110. However, allowing only five binary digits, this infinite number will be truncated to 0.0110. Such truncation introduces a round off error. This would be true for *any* precision. This is analogous to representing a rational number such as $\frac{1}{3} = 0.3333, \ldots$ in the decimal system. See equations (11) and (13).

**9**. For numerous examples in various economic settings, again see Day (1994, 2000).

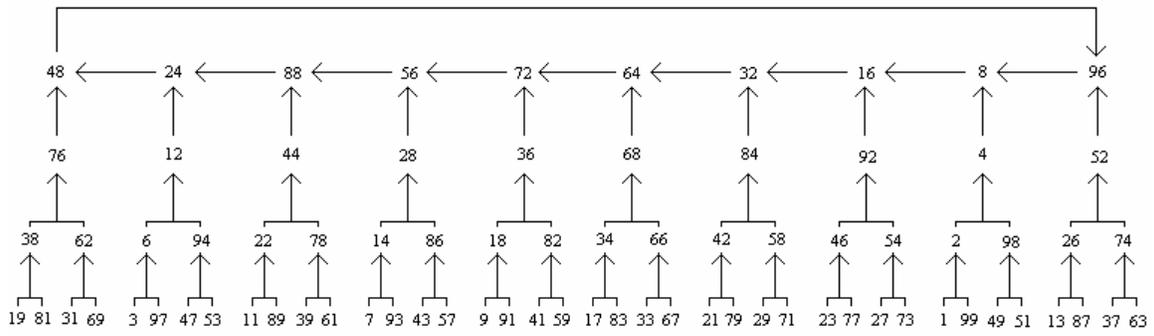

Figure 1

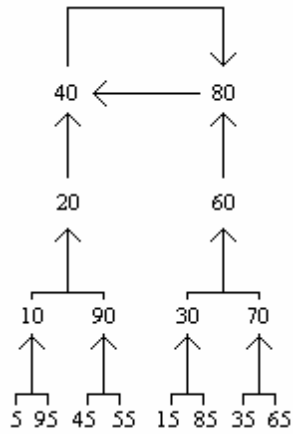

Figure 2

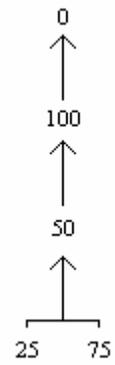

Figure 3

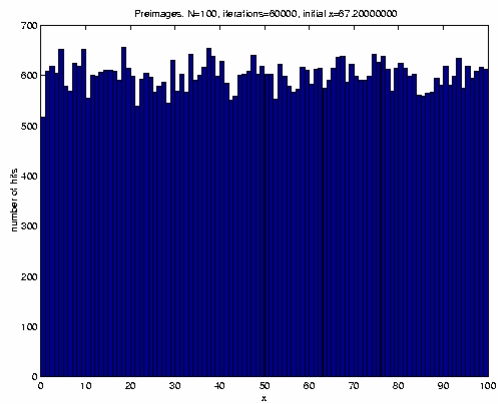

Figure 4

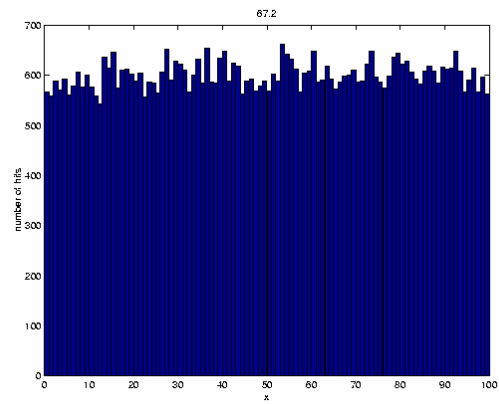

Figure 5